\documentclass[conference]{IEEEtran}
\IEEEoverridecommandlockouts
\usepackage{cite}
\usepackage{amsmath,amssymb,amsfonts,mathtools}
\usepackage{stfloats}
\usepackage{algorithm}
\usepackage{algorithmic}

\usepackage{pgfplots}
 \pgfplotsset{compat=1.14}

\usepackage{subcaption}
\usepackage{multirow}

\usepackage[numbers]{natbib}

\usepackage{txfonts}
\usepackage{times}
\newcommand{\ftextnumero}{{\fontfamily{txr}\selectfont \textnumero}}

\usepackage{tablefootnote}

\usepackage{natbib}
\usepackage{caption}

\makeatletter
\newcommand{\newlineauthors}{%
  \end{@IEEEauthorhalign}\hfill\mbox{}\par
  \mbox{}\hfill\begin{@IEEEauthorhalign}
}
\makeatother

\usepackage{graphicx}
\usepackage{textcomp}
\usepackage{xcolor}
\usepackage{lipsum}
\def\BibTeX{{\rm B\kern-.05em{\sc i\kern-.025em b}\kern-.08em
    T\kern-.1667em\lower.7ex\hbox{E}\kern-.125emX}}
\begin{document}

\title{Hardware Implementation of Soft Mapper/Demappers in Iterative EP-based Receivers
}

\author{
    Ian Fischer Schilling\textsuperscript{*}, 
    Serdar \c{S}ahin\textsuperscript{\dag}, 
    Camille Leroux\textsuperscript{*}, 
    Antonio Maria Cipriano\textsuperscript{\dag}, 
    Christophe J\'{e}go\textsuperscript{*}
\newlineauthors
\and
\IEEEauthorblockA{\textsuperscript{*}\textit{University of Bordeaux, Bordeaux INP}\\
IMS Lab, UMR CNRS 5218, France \\
firstname.last-name@ims-bordeaux.fr}
\and
\IEEEauthorblockA{\textsuperscript{\dag}
\textit{Thales}\\
Gennevilliers, France \\
firstname.last-name@thalesgroup.com}
\thanks{This work has been funded by the French National Research Agency under grant number ANR-20-CE25-0008-01 (EVASION Project: https://anr-evasion.ims-bordeaux.fr/).}
}

\maketitle

\begin{abstract}
This paper presents a comprehensive study and implementations onto FPGA device of an Expectation Propagation (EP)-based receiver for QPSK, 8-PSK, and 16-QAM. To the best of our knowledge, this is the first for this kind of receiver. The receiver implements a Frequency Domain (FD) Self-Iterated Linear Equalizer (SILE), where EP is used to approximate the true posterior distribution of the transmitted symbols with a simpler distribution. Analytical approximations for the EP feedback generation process and the three constellations are applied to lessen the complexity of the soft mapper/demapper architectures. The simulation results demonstrate that the fixed-point version performs comparably to the floating-point. Moreover, implementation results show the efficiency in terms of FPGA resource usage of the proposed architecture.
\end{abstract}


\section{Introduction}

In digital communication systems, achieving minimal error rates in data detection and/or decoding requires the resolution of a Maximum A Posteriori (MAP) or Maximum Likelihood (ML) problem \cite{Bahl74}. However, the computational complexity of resolving such criteria is often prohibitive, particularly in real-world frequency selective channels, where the number of computations increases exponentially with factors such as data length, modulation order, and channel memory. As a result, practical receiver design often involves applying simplifying hypotheses and approximations. One promising approach in the context of Frequency Domain (FD) Linear Equalization (LE) is equalizers designed with Expectation Propagation (EP). Indeed, they have demonstrated an appealing trade-off between performance and computational complexity \cite{Sahin18}.

In this paper, a comprehensive study and implementation of an EP-based receiver for communications over frequency selective channels using standard Phase Shift Keying (PSK) or Quadrature Amplitude Modulation (QAM) constellations is presented. 
The receiver implements an FD Self-Iterated Linear Equalizer (SILE), where EP is applied to approximate the true posterior distribution of the transmitted symbols with a simpler distribution that can be easily manipulated. 
Previously, the implementation of a simplified EP receiver for multiple antenna receivers has been reported in \cite{Auras14}. 
A low complexity EP detector for sparse code multiple access was also proposed in \cite{Xiao19}. However, the authors only study the impact of simplifications on the performance and estimate the computational cost of their proposal per operation type.
Simplified EP-based FD equalization is studied in \cite{Cipriano23}. Similar to the previous case, only a computational complexity assessment was provided.
To the best of our knowledge, we propose the first hardware implementation of an EP-based FD SILE receiver.

The contributions of this paper are the following:
\begin{enumerate}
\item Methods to reduce the complexity of the EP-based FD-SILE algorithm are proposed, which are different from the ones in \cite{Cipriano23}. They include a new way to generate the EP soft feedback and also a new method to calculate extrinsic Log-Likelihood Ratios (LLR) for 8-PSK.
\item A fixed-point version of the model enables to verify that the degradation in terms of performance due to these new algorithmic simplifications is limited.
\item The implementation of soft mapper/demapper architectures is carried out on a Field Programmable Gate Array (FPGA), specifically the PYNQ Z2 board. 
\end{enumerate}

 The PYNQ Z2 board contains a device that combines an ARM processor and an FPGA, which enables easier Ethernet communication. The implementation was done in a Hardware in the Loop (HIL)  configuration, with the EP parts implemented onto the FPGA, while the others run on a computer thanks to py-AFF3CT, a Python wrapper for the Forward Error Correction Toolbox AFF3CT \cite{Cassagne19}. The analysis of FPGA resource usage shows very low complexity overhead for three different and widely used constellations. These results confirm that the proposed EP equalizer is potentially a good candidate for practical implementation even on cost- and complexity-constrained digital communication equipment.

The paper is organized as follows. 
A description of the FD SILE receiver with EP is provided in Section~\ref{SILE}. The simplifications and analytical approximations applied to the soft mapper/demapper to decrease its 
complexity are presented in Section~\ref{Approx}. The fixed-point conversion analysis to facilitate the soft mapper/demapper architecture design is presented in Section~\ref{FPconv}. The architecture implementation and FPGA prototyping, with the experimental setup and resource usage, are detailed in Section~\ref{Res}. The paper is concluded in Section~\ref{Concl}.

\section{EP-Based FD-SILE Algorithm}
\label{SILE}

\begin{figure}[tbp]
	\centering
    \includegraphics[width=0.49\textwidth]{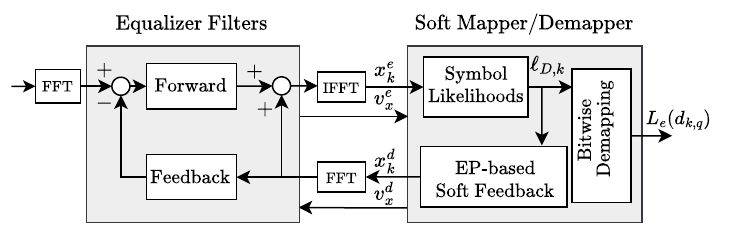}
	\caption{FD SILE functional structure.}
	\label{fig_fdleic}
\label{fig:archi}
\end{figure}

Expectation Propagation (EP) is a powerful technique exploited in statistical inference for approximating complex probability distributions by simpler distributions from the exponential family through moment matching. The FD self-iterated linear equalizer (SILE) algorithm derived in \cite{Sahin18} is based on EP to compute extrinsic soft decision feedback. 

The functional structure of the iterative receiver is illustrated in Fig.~\ref{fig_fdleic}. 
The received signal is first mapped to the frequency domain thanks to an FFT function. Then, a linear Minimum Mean Square Error (MMSE) filter with interference cancellation is used for equalization.
In addition to the channel frequency response and noise statistics, the filter computation requires the statistics of the soft decision feedback \cite{Sahin18}. 

The receiver performs $S$ self-iterations which go through the equalizer, the soft demapper, and the EP-based soft mapper. 
At self-iteration $s=0,\dots,S$, after the filtering stage, the equalized symbols in the time-domain $x_k^{e(s)}$ are fed into the soft demapper, along with an estimate of the residual post-equalization noise and interference variance $v_x^{e(s)}$.
The soft demapper then estimates the unnormalized log-likelihood distribution $\ell_{D,k}^{(s)}(\alpha)$ with
\begin{equation} \label{log_d}
    \ell_{D,k}^{(s)}(\alpha)
    = -|x_k^{e(s)} - \alpha|^2 / v_x^{e(s)}, \forall{k},  \forall{\alpha} \in X,
\end{equation}
where $X$ is the symbol constellation and $K$ is the equalized block length.

If $s < S$, the EP soft mapper computes the feedback for the next iteration by first computing the normalized a posteriori distribution in the linear domain, with $\forall{k},  \forall{\alpha} \in X$,
\begin{equation} \label{d_norm}
    D_k^{(s)}(\alpha) = \exp{(\ell_{D,k}^{(s)}(\alpha))} / \textstyle\sum_{\alpha' \in X} \exp({\ell_{D,k}^{(s)}(\alpha'))}.
\end{equation}
Next, the soft a posteriori symbol estimates $\mu_k^{d(s)}$ and variances $\gamma_x^{d(s)}$ can be computed with the moments of $D_k^{(s)}$:
\begin{align}
     \mu_k^{d(s)} =& \textstyle\sum_{\alpha \in X} \alpha D_k^{(s)}(\alpha), \forall{k}, \label{mu_se}\\
     \gamma_{x, k}^{d(s)} = & \textstyle\sum_{\alpha \in X} |\alpha|^2 D_k^{(s)}(\alpha) - |\mu_k^{d(s)}|^2, \forall{k}, \label{gamma_se1}\\
     \gamma_x^{d(s)} =& K^{-1} \sum\nolimits_{k=1}^{K} \gamma_{x, k}^{d(s)}, \label{gamma_se2}
\end{align}
The average variance $\gamma_x^{d(s)}$ enables the proper use of EP for FD equalization.
Finally, the extrinsic soft feedbacks based on EP are obtained by performing a Gaussian PDF division:
\begin{equation} \label{pdf_var}
    \small 
    x_k^{d(s+1)} = \frac{\mu_k^{d(s)} v_x^{e(s)} - x_k^{e(s)} \gamma_x^{d(s)}}{v_x^{e(s)} - \gamma_x^{d(s)}}, \,
    v_x^{d(s+1)} = \frac{v_x^{e(s)} \gamma_x^{d(s)}}{v_x^{e(s)} - \gamma_x^{d(s)}}.
\end{equation}

Exponential smoothing could be applied across self-iterations, as done in \cite{Cipriano23}, for stabilizing convergence. But, in order to simplify this initial work, it is not considered. 

At the last self-iteration $s=S$, only bit-wise extrinsic LLRs for soft decoding  are computed \cite{Cipriano23}:
\begin{equation} \label{ext_llr}
    L_e(d_{k,q}) =  \textstyle \mathrm{log}\sum_{\alpha \in  X_q^0}e^{\ell_{D,k}^{(s)}(\alpha)} - \mathrm{log}\sum_{\alpha' \in X_q^1}e^{\ell_{D,k}^{(s)}(\alpha')}.
\end{equation}

It is important to note that the FD SILE algorithm is considerably less computationally intensive than alternative iterative equalizers in the time domain \cite{Sahin18}.
Nevertheless, logarithmic and exponential operations within soft mapping and demapping still pose a significant implementation challenge and scale poorly with the constellation size. 
To illustrate this point, the number of floating-point operations (FLOP) involved in the exact equalization, soft mapping, and soft demapping are provided in Table~\ref{tab_complex}. These numbers mostly depend on the constellation size and the number of self-iterations. 
They are obtained through the same approach as in \cite{Sahin18}, where operations such as addition, multiplication, and division operation are counted with weights.

\section{Analytical Approximations}\label{Approx}

In this section, we discuss further simplifications of the complexity of EP-based soft feedback computation for practical constellations of QPSK, 8-PSK, and 16-QAM.

\subsection{Simplified EP-based feedback computation}
\label{simp_feed_comp}

\begin{figure}[tbp]
\centering
\begin{subfigure}{0.32\columnwidth}
    \includegraphics[width=\columnwidth]{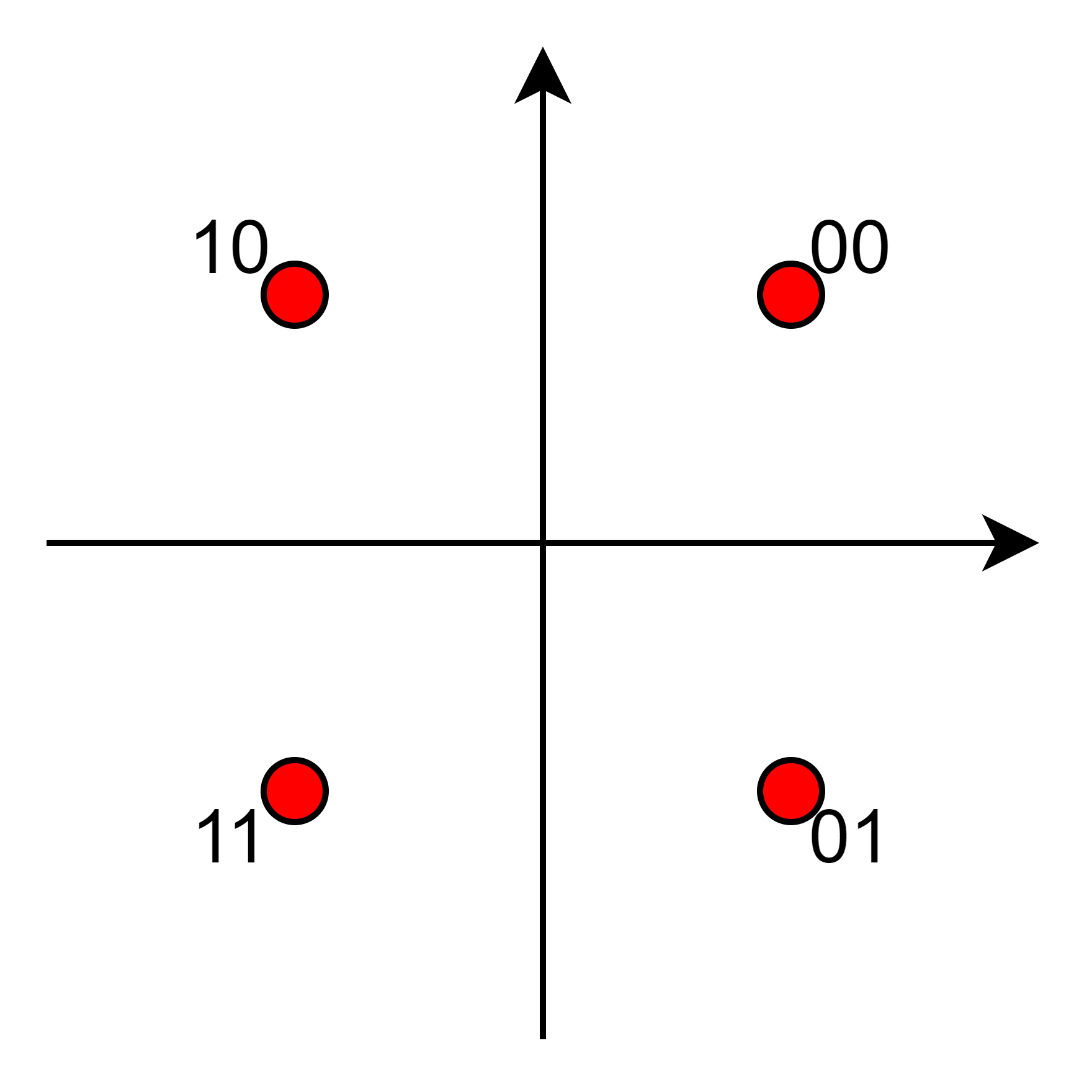}
    \caption{QPSK.}
    \label{fig:qpsk}
\end{subfigure}
\hfill
\begin{subfigure}{0.32\columnwidth}
    \includegraphics[width=\columnwidth]{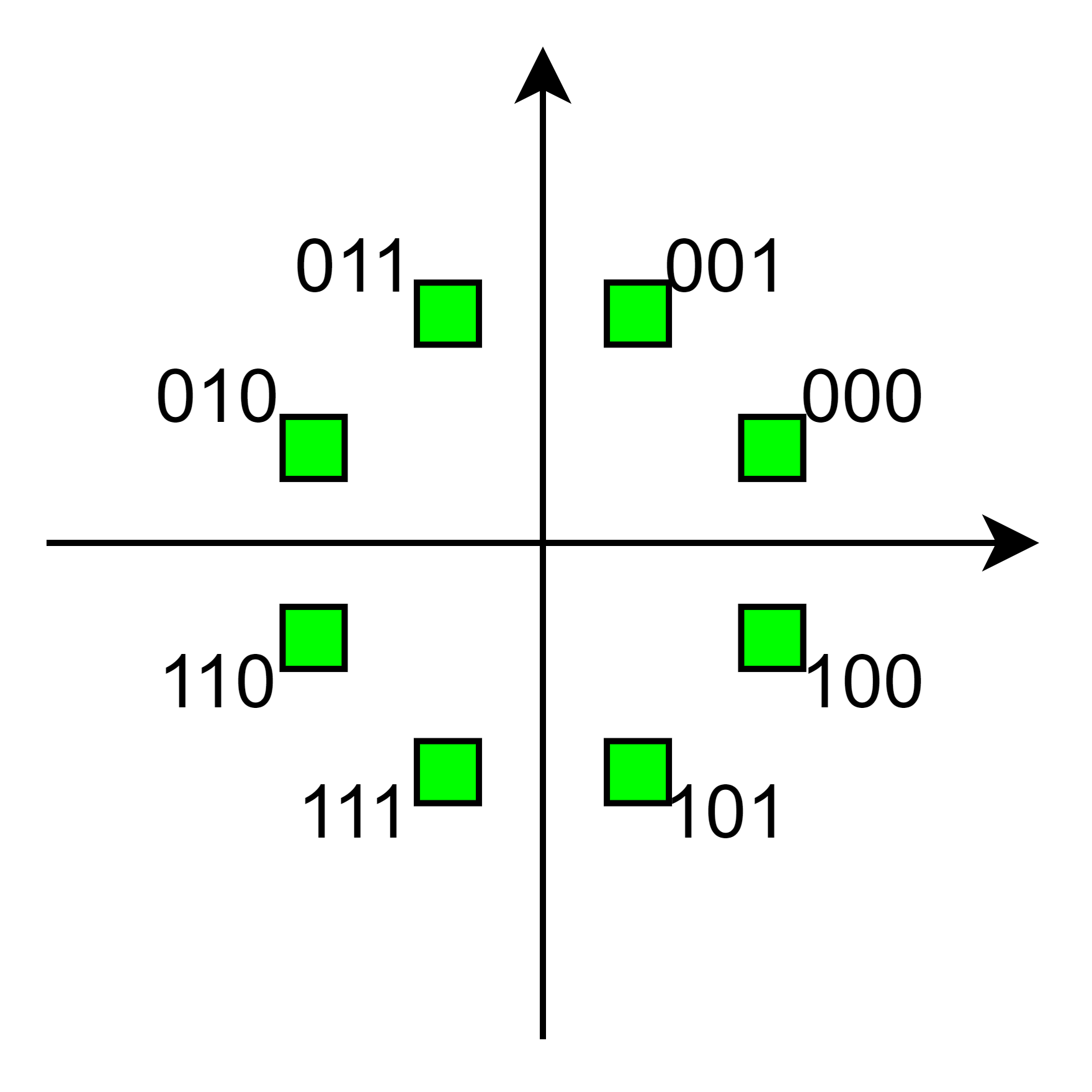}
    \caption{8-PSK.}
    \label{fig:8psk}
\end{subfigure}
\hfill
\begin{subfigure}{0.32\columnwidth}
    \includegraphics[width=\columnwidth]{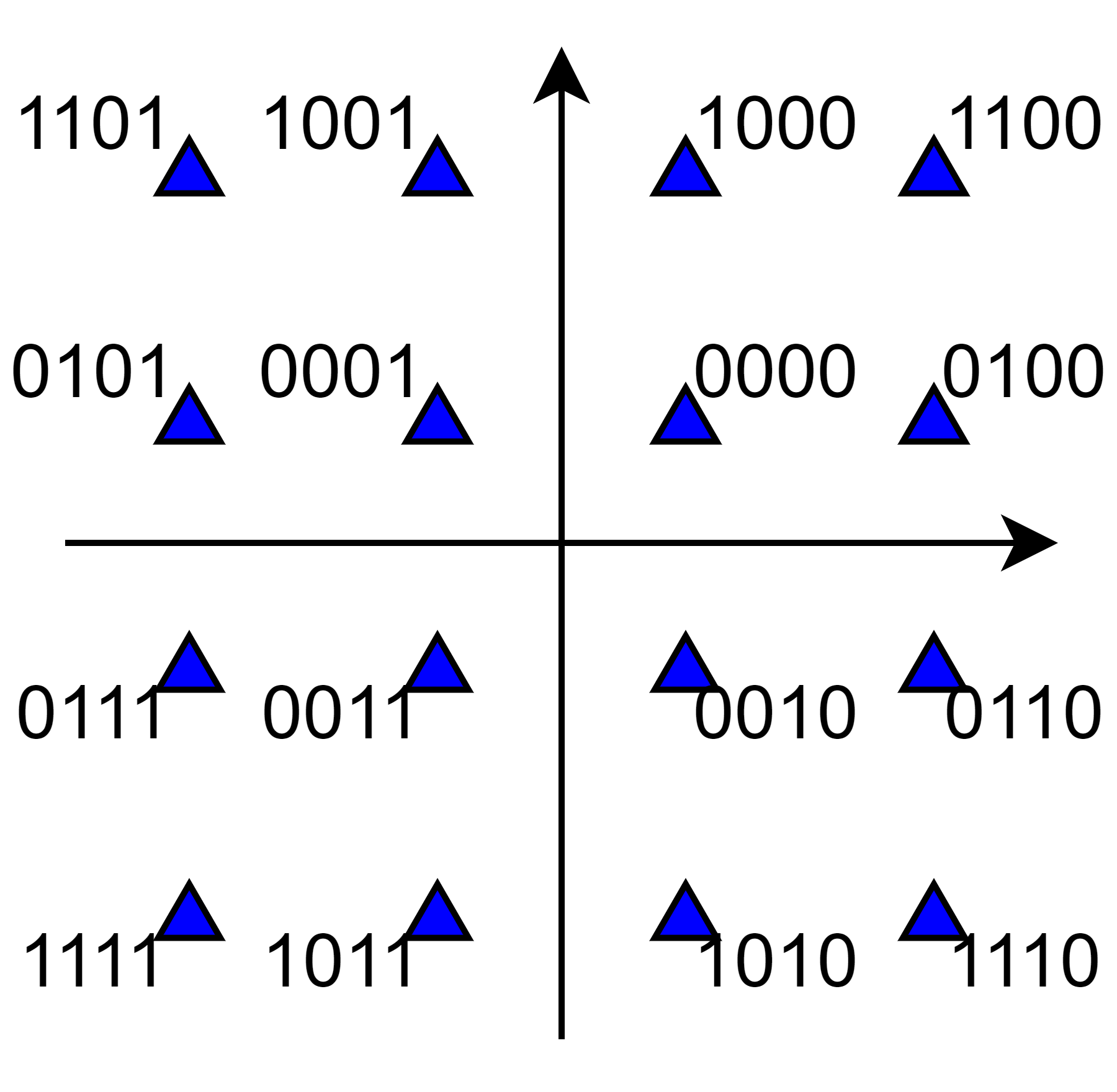}
    \caption{16-QAM.}
    \label{fig:16qam}
\end{subfigure}
        
\caption{Constellation bit values.}
\label{fig:bit_val}
\end{figure}

In \cite{Cipriano23}, the use of the asymptotic a posteriori mean-square error (MSE) $\tilde{\gamma}_x^{d}$, as a function of $v_x^e$, instead of $\gamma_x^{d}$, is shown to reduce soft mapping complexity. Moreover, it also improves receiver robustness. 
In particular, when tabulating the auxiliary quantity $C_{EP}(v_x^e) = \tilde{\gamma}_x^d/(v_x^e - \tilde{\gamma}_x^d)$, the computational complexity can be further reduced through 
\begin{align}
  v_x^{d(s+1)} =&\, v_x^{e(s)} C_{EP}(v_x^{e(s)}) \label{var_cep},\\
  x_k^{d(s+1)} =&\, \mu_k^{d(s)} + C_{EP}(v_x^{e(s)})(\mu_k^{d(s)} - x_k^{e(s)}), \forall{k}. \label{mean_cep} 
\end{align}
Although this alleviates part of the soft mapping complexity issue, the computation of a posteriori estimates $\mu_k^{d}$ and extrinsic LLRs $L_e(d_{k,q})$ still remains computationally intensive.

\subsection{Simplified soft demapping}
\label{simp_soft_demap}

\begin{figure}[tbp]
	\centering
    \includegraphics[width=0.35\textwidth]{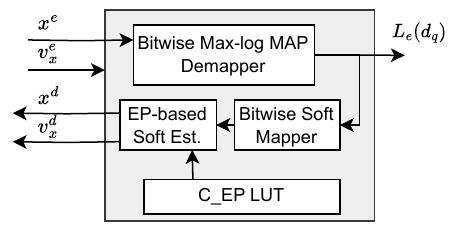}
	\caption{Simplified bitwise soft demapper.}
	\label{fig_bitwiseDemap}
\end{figure}

In \cite{Cipriano23}, piece-wise linear approximations of $\mu_k^{d(s)}$ as a function of $x_k^{e(s)}$ were proposed to lower the computational complexity of $\mu_k^{d(s)}$ with small performance degradation. 
However, this approach is limited to square QAM constellations, and it is too complex for turbo-equalization.
In this work, a \textit{new simplified computation of EP-based feedback is proposed} based on bitwise soft demapping, followed by bitwise soft mapping, for computing both $L_e(d_{q})$ and $\mu^{d}$. 
This structure is illustrated in Fig.~\ref{fig_bitwiseDemap}. The symbol index $k$ and the iteration index $s$ have been dropped for the sake of readability.

\begin{table}
    \centering
    \begin{tabular}{cc|c|cc|cc}
       \multirow{2}{2.2em}{Const.} & \multirow{2}{1em}{$S$}  & \multirow{2}{2.2em}{Equ.} & \multicolumn{2}{c}{Exact} & \multicolumn{2}{c}{Simplified} \\
        &  &  & Demap. & Map. & Demap. & Map. \\      
        \hline
       \multirow{3}{3em}{QPSK} & 0 & 74 & 58 & 0 & 3 & 0  \\
       & 1 & 153 & 58 & 105 & 6 & 26  \\
       & 2 & 231 & 58 & 210 & 9 & 53  \\
       \hline
       \multirow{3}{3em}{8-PSK} & 0 & 74 & 195 & 0 & 12 & 0 \\
       & 1 & 153 & 195 & 185 & 24 & 33  \\
       & 2 & 231 & 195 & 370 & 36 & 67 \\
       \hline
       \multirow{3}{3em}{16-QAM}& 0 & 74 & 548 & 0 & 14 & 0 \\
       & 1 & 153 & 548 & 345 & 28 & 40 \\
       & 2 & 231 & 548 & 690 & 42 & 81
    \end{tabular}
    
    \caption{Computational complexity of components of FD SILE in FLOP/symbol.}
    \label{tab_complex}
\end{table}

A widespread means of carrying out soft demapping with reasonable complexity is through the simplification of 
the log-sum-exp, twice used in eq. (\ref{ext_llr}). Indeed, one can replace this function with a maximum (max-log-MAP) \cite{Erfanian94}, and remains optimal for the QPSK constellation. However, some performance loss can be expected for higher-order constellations.

Besides, the max-log-MAP demapper remains prohibitive for high-order constellations. With $M$ being the constellation size, for each symbol, it requires computing $M$ squared-distances and performing $M$ comparisons for each of the $Q=\mathrm{log}_2 M$ bits.
Various proposals in the literature provide direct linear approximations of max-log-MAP LLRs $L_e(d_{q})$ as a function of $x^e$ and $v_x^e$ for each $q$, computed through the geometric properties of each constellation.
Indeed, some practical implementations in the literature \cite{Ali15} rely on analytical expressions of the extrinsic max-log MAP LLRs. In particular, there are closed-form analytical equations for Gray-mapped QAM constellations \cite{Tosato02}. 
Proposed expressions for the QPSK and 16-QAM constellations are provided in Table~\ref{tab_softDemap}.

\begin{table}
    \centering
    \begin{tabular}{|c|l|}
    \hline
       Constellation  & Bitwise Log-Likelihood Ratios \\
       \hline 
        QPSK & $L_e(d_1) = 2\sqrt{2}R(x^e)/v_x^e$, \,  $L_e(d_2) = 2\sqrt{2}I(x^e)/v_x^e$  \\
       \hline 
       \multirow{9}{4em}{16-QAM}  & $L_e(d_1) = 4d(2d - |I(x^e)|)/v_x^e$\\
            & $L_e(d_2) = 4d(2d - |R(x^e)|)/v_x^e$\\
            & $L_e(d_3) = 
            \begin{dcases}
                4dI(x^e)/v_x^e,& |I(x^e)|<2d\\
                8d(I(x^e)-d)/v_x^e,& I(x^e) > 2d\\
                8d(I(x^e)+d)/v_x^e,& I(x^e) < -2d
            \end{dcases}$\\
            & $L_e(d_4) = 
            \begin{dcases}
                4dR(x^e)/v_x^e,& |R(x^e)|<2d\\
                8d(R(x^e)-d)/v_x^e,& R(x^e) > 2d\\
                8d(R(x^e)+d)/v_x^e,& R(x^e) < -2d
            \end{dcases}$\\
            \hline
    \end{tabular}
    
    \caption{Bitwise soft demapping expressions.}
    \label{tab_softDemap}
\end{table}

\begin{table}
    \centering
    \begin{tabular}{c|lll}
       $\text{LUT}_{\text{8PSK}}$  & $\Delta_{\alpha^*,1}$ & $\Delta_{\alpha^*,2}$ & $\Delta_{\alpha^*,3}$ \\
       \hline
       $\alpha_{0}$ & 1.0824 - 1.0824j  & 2.6131 - 1.0824j &  0.0000 + 1.5307j  \\
       $\alpha_{1}$ & 1.0824 - 1.0824j  & 1.5307 - 0.0000j & -1.0824 + 2.6131j \\
       $\alpha_{2}$ & -1.0824 - 1.0824j  & 2.6131 + 1.0824j &  0.0000 + 1.5307j \\
       $\alpha_{3}$ & -1.0824 - 1.0824j  & 1.5307 - 0.0000j  & 1.0824 + 2.6131j \\
       $\alpha_{4}$ & 1.0824 + 1.0824j  & 2.6131 + 1.0824j  & 0.0000 + 1.5307j \\
       $\alpha_{5}$ & 1.0824 + 1.0824j &  1.5307 - 0.0000j  & 1.0824 + 2.6131j \\
       $\alpha_{6}$ & -1.0824 + 1.0824j &  2.6131 - 1.0824j &  0.0000 + 1.5307j \\
       $\alpha_{7}$ & -1.0824 + 1.0824j  & 1.5307 - 0.0000j & -1.0824 + 2.6131
    \end{tabular}
    
    \caption{Look-up-table content for 8-PSK demapping.}
    \label{tab_lut8psk}
\end{table}

But regarding non-square constellations, as the 8-PSK, the geometric characterization of max-log MAP LLRs does not systematically have closed-form expressions. 
For such Gray-mapped PSK constellations, \cite{Wang14} exploits labeling properties of the constellation, and the hard decision on the equalized symbol to compute only two distances per LLR.
This method performs exact max-log-MAP by exploiting the properties of Gray labeling for finding the likely closest symbol from the set opposite to the hard decision. However, the arithmetic computation of the opposite symbol and the computation of the formula with Euclidean distances still have a significant weight within the overall computational complexity.

Here we propose to further simplify the computation of 8-PSK LLRs in \cite{Wang14}, by applying a semi-analytical approach \cite{Sahin_brev}.
Indeed, the max-log-MAP LLR is given by
\begin{equation}
    L_e(d_{q}) = -(1-2\hat{d}_{q}^*) ({|x^e-\alpha^*|^2 -|x^e-\alpha_{\Bar{q}}^*|^2})/{v_x^e},
\end{equation}
where $\alpha^*$ is the closest constellation point to $x^e$, with $q^{th}$ bit of the $\alpha^*$'s label being denoted $\hat{d}_{q}^*$ . Furthermore, $\alpha_{\Bar{q}}^*$ is the constellation point that corresponds to the closest symbol to $\alpha^*$ that has the opposite value on the $q^{th}$ bit.

Instead of computing $\alpha_{\Bar{q}}^*, \forall q$ with $Q^2$ additions as in \cite{Wang14}, we rewrite the expression of the max-log-MAP LLR as 
\begin{equation} \label{llr_8psk}
    L_e(d_q) = ({R(x^e) R(\Delta_{\alpha^*, q}) + I(x^e) I(\Delta_{\alpha^*, q})})/{v_x^e},
\end{equation}
where $\Delta_{\alpha^*, q} = 2(1-2\hat{d}_{q}^*)(\alpha^*-\alpha_{\Bar{q}}^*)$. 
Hence, $\Delta_{\alpha^*, q}$ can be precomputed and stored in a LUT, as shown in Table \ref{tab_lut8psk}, for each $\alpha^* \in X$ and $q=1,\dots,Q$:
\begin{equation}
    (\Delta_{\alpha, 1}, \Delta_{\alpha, 2}, \Delta_{\alpha, 3}) = \text{LUT}_{\text{8PSK}}(\alpha), \alpha\in X.
\end{equation}
To access the LUT, a hard decision has to be made through comparisons on $x^e$, to compute the symbol label $m$ in decimal:
\begin{equation} \label{hard_dec}
    \small m = 4(I(x^e) < 0) +2(R(x^e) < 0) + (|R(x^e)| < |I(x^e)|).
\end{equation}

\subsection{Simplified soft mapping}

\begin{table}
    \centering
    \begin{tabular}{|c|l l|}
    \hline
       Constel.  & \multicolumn{2}{ c| }{Soft Estimates} \\\hline
       QPSK  & $R(\mu^{d}) = p_1/\sqrt{2}$ &
                $I(\mu^{d}) = p_2/\sqrt{2}$ \\\hline
       8-PSK  & $R(\mu^{d}) = (b_8 + a_8 p_1) p_2$ &
                 $I(\mu^{d}) = (b_8 - a_8 p_1) p_3$\\\hline
       16-QAM  & $R(\mu^{d}) = (2-p_2)p_4/\sqrt{10}$ &
                 $I(\mu^{d}) = (2-p_1)p_3/\sqrt{10}$\\\hline\hline
        {Params.} &\multicolumn{2}{ l| }{\small{$a_8=\sqrt{(2-\sqrt{2})/8},\,  b_8=\sqrt{(2+\sqrt{2})/8}$}} \\
        \hline
    \end{tabular}
    
    \caption{Bitwise soft mapping expressions.}
    \label{tab_softMap}
\end{table}

To simplify the computation of the soft symbol estimate $\mu^{d}$ from (\ref{mu_se}), analytical bitwise soft mapping can be applied.
This enables to take into account probabilities on symbols through bitwise LLRs $L(d_q)$ and also to compute $\mu^{d}$  without an intermediate step. 
Thanks to Gray mapping, assuming that bits within a symbol are independent, we obtain 
\begin{equation}
 \textstyle\mathbb{P}(x=\alpha) = \prod_{q=1}^Q \mathbb{P}(d_q = \phi^{-1}_q(\alpha)),
\end{equation}
where $\phi^{-1}_q(\cdot)$ yields the $q^\text{th}$ bit of $\alpha \in X$. In addition, as $\mathbb{P}(d_q = b)=(1+(1-2b)p_q)/2$, where $p_q=\tanh(L(d_q))/2)$, one can exploit the geometry of the constellations with respect to real and imaginary parts and the labeling, to derive expressions for $\mu_k^{d(s)}$ as a function of soft bits $p_q$.
Note that, in our case, LLRs for soft mapping are extrinsic $L(d_q)= L_e(d_q)$. Nevertheless, this approach can be readily generalized to a turbo-equalizer if a priori LLRs $L_a(d_q)$ from a soft output decoder are available with $L(d_q) = L_e(d_q) + L_a(d_q)$.

Such soft mapping techniques have been previously applied in \cite{Tomasoni06} and \cite{Auras14} for APP-based and EP-based soft feedback, respectively. However, in this study, these formulas are only required for the mean of soft APP estimates, as the EP-based variance is handled through the tabulation technique of \cite{Cipriano23}, with $C_{EP}$ as stated earlier in Section \ref{simp_feed_comp}.

In addition, for the mean estimates, the computation of the hyperbolic tangent could be further simplified, by a piece-wise linear approximation, where all the slope and bias coefficients are powers of two. This enables a more efficient fixed-point implementation, as expressed below:
\begin{equation}
    \small \mathrm{tanh}(x) =
    \begin{dcases}
        x,& |x|<0.5,\\
        0.5x +0.25\mathrm{sign}(x),& 0.5\leq |x| < 1.0,\\
        0.25x +0.5\mathrm{sign}(x),& 1.0\leq |x| < 2.0,\\
        \mathrm{sign}(x),& \text{otherwise}.
    \end{dcases}
\end{equation}

The complete expressions of soft estimates for constellations of interest are provided in Table~\ref{tab_softMap}. This concludes the algorithmic simplifications in soft mapping and demapping, and the resulting algorithmic complexity is given in Table~\ref{tab_complex}.

\section{Fixed-Point Conversion Analysis}
\label{FPconv}

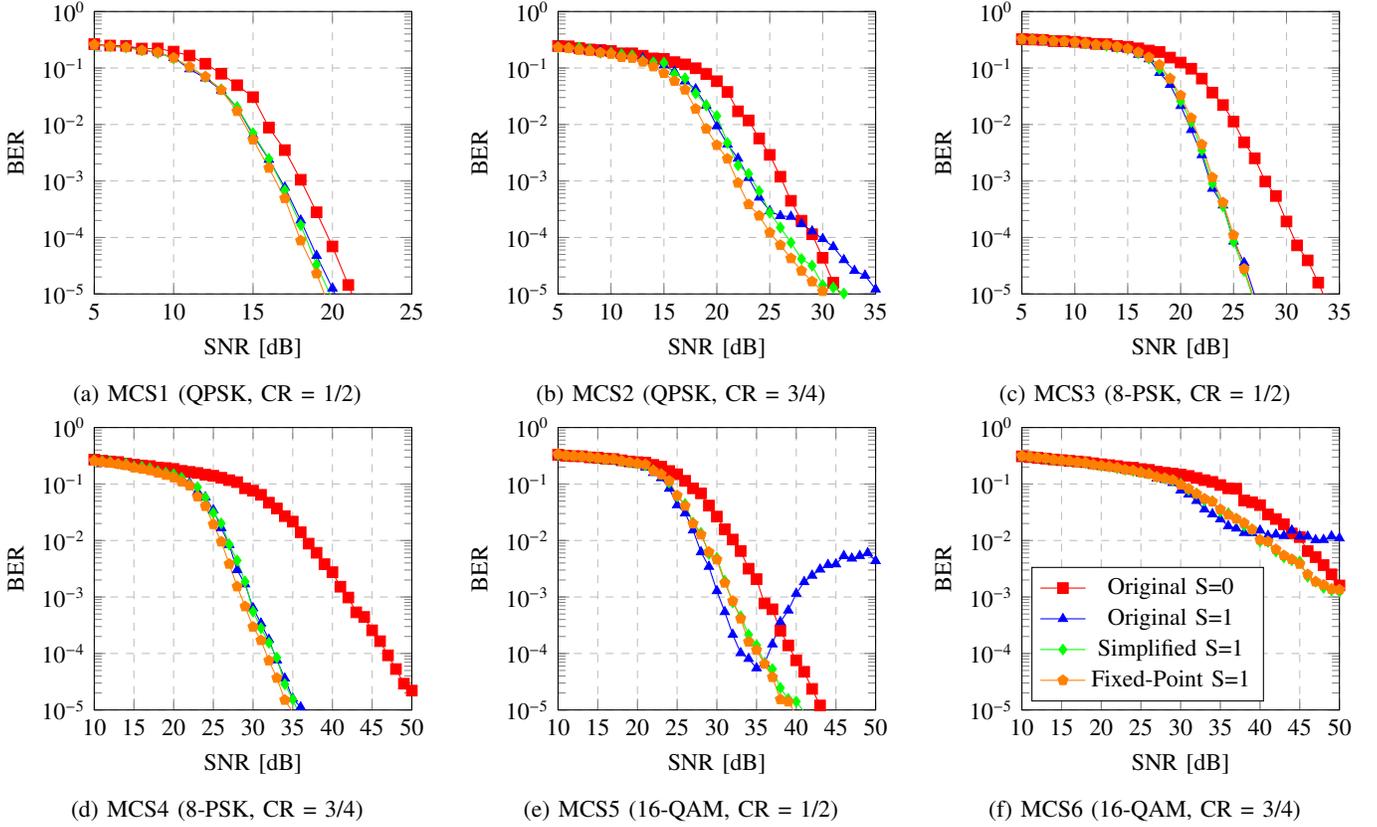
\begin{figure*}[t]
\centering
\begin{subfigure}[h]{0.32\textwidth}
\vspace{-0.4cm}
\begin{tikzpicture}[font=\small]
\begin{axis}[
    width=1\linewidth,
    height=0.92\linewidth,
    ymode=log,
    xlabel={SNR [dB]},
    ylabel={BER},
    xmin=5, xmax=25,
    ymin=1e-5, ymax=1,
    xtick={0,5, 10, 15, 20, 25},
    legend pos=south west,
    xmajorgrids=true,
    ymajorgrids=true,
    grid style=dashed,
]

\addplot[
    color=red,
    mark=square*,
    ]
    coordinates {
    (0, 3.130000e-01)
    (1, 2.980000e-01)
    (2, 2.970000e-01)
    (3, 2.920000e-01)
    (4, 2.800000e-01)
    (5, 2.650000e-01)
    (6, 2.500000e-01)
    (7, 2.470000e-01)
    (8, 2.210000e-01)
    (9, 2.220000e-01)
    (10, 1.950000e-01)
    (11, 1.660000e-01)
    (12, 1.190000e-01)
    (13, 7.920000e-02)
    (14, 4.960000e-02)
    (15, 3.050000e-02)
    (16, 8.810000e-03)
    (17, 3.520000e-03)
    (18, 1.050000e-03)
    (19, 2.790000e-04)
    (20, 6.930000e-05)
    (21, 1.440000e-05)
    (22, 3.030000e-06)
    };

\addplot[
    color=blue,
    mark=triangle*,
    ]
    coordinates {
    (0, 3.140000e-01)
    (1, 2.970000e-01)
    (2, 2.950000e-01)
    (3, 2.930000e-01)
    (4, 2.820000e-01)
    (5, 2.560000e-01)
    (6, 2.420000e-01)
    (7, 2.390000e-01)
    (8, 2.110000e-01)
    (9, 1.880000e-01)
    (10, 1.520000e-01)
    (11, 9.580000e-02)
    (12, 6.600000e-02)
    (13, 4.030000e-02)
    (14, 2.000000e-02)
    (15, 6.500000e-03)
    (16, 2.390000e-03)
    (17, 7.690000e-04)
    (18, 2.010000e-04)
    (19, 4.760000e-05)
    (20, 1.260000e-05)
    (21, 2.670000e-06)
    (22, 0)
    };
    
\addplot[
    color=green,
    mark=diamond*,
    ]
    coordinates {
    (0, 3.070000e-01)
    (1, 2.900000e-01)
    (2, 2.940000e-01)
    (3, 2.920000e-01)
    (4, 2.780000e-01)
    (5, 2.610000e-01)
    (6, 2.420000e-01)
    (7, 2.320000e-01)
    (8, 2.060000e-01)
    (9, 1.840000e-01)
    (10, 1.480000e-01)
    (11, 1.050000e-01)
    (12, 6.940000e-02)
    (13, 4.200000e-02)
    (14, 2.020000e-02)
    (15, 6.960000e-03)
    (16, 2.450000e-03)
    (17, 6.870000e-04)
    (18, 1.660000e-04)
    (19, 3.310000e-05)
    (20, 7.860000e-06)
    (21, 0)
    (22, 0)
    };

\addplot[
    color=orange,
    mark=pentagon*,
    ]
    coordinates {
    (0, 3.150000e-01)
    (1, 2.960000e-01)
    (2, 3.000000e-01)
    (3, 2.910000e-01)
    (4, 2.710000e-01)
    (5, 2.580000e-01)
    (6, 2.420000e-01)
    (7, 2.320000e-01)
    (8, 2.050000e-01)
    (9, 1.890000e-01)
    (10, 1.510000e-01)
    (11, 1.050000e-01)
    (12, 7.020000e-02)
    (13, 4.070000e-02)
    (14, 1.720000e-02)
    (15, 5.400000e-03)
    (16, 1.700000e-03)
    (17, 4.940000e-04)
    (18, 8.800000e-05)
    (19, 2.300000e-05)
    (20, 4.190000e-06)
    (21, 0)
    (22, 0)
    };
    
\end{axis}
\end{tikzpicture}
\captionsetup{justification=centering}
\caption{\small MCS1 (QPSK, CR = 1/2)}
\end{subfigure}
\hfill
\begin{subfigure}[h]{0.32\textwidth}
\vspace{-0.4cm}
\begin{tikzpicture}[font=\small]
\begin{axis}[
    width=1\linewidth,
    height=0.92\linewidth,
    ymode=log,
    xlabel={SNR [dB]},
    ylabel={BER},
    xmin=5, xmax=35,
    ymin=1e-5, ymax=1,
    xtick={0,5, 10, 15, 20, 25, 30, 35},
    legend pos=south west,
    xmajorgrids=true,
    ymajorgrids=true,
    grid style=dashed,
]

\addplot[
    color=red,
    mark=square*,
    ]
    coordinates {
    (0, 2.920000e-01)
    (1, 2.810000e-01)
    (2, 2.680000e-01)
    (3, 2.670000e-01)
    (4, 2.520000e-01)
    (5, 2.450000e-01)
    (6, 2.420000e-01)
    (7, 2.250000e-01)
    (8, 2.130000e-01)
    (9, 2.080000e-01)
    (10, 2.000000e-01)
    (11, 1.820000e-01)
    (12, 1.820000e-01)
    (13, 1.650000e-01)
    (14, 1.480000e-01)
    (15, 1.450000e-01)
    (16, 1.270000e-01)
    (17, 1.160000e-01)
    (18, 9.910000e-02)
    (19, 7.830000e-02)
    (20, 5.840000e-02)
    (21, 3.770000e-02)
    (22, 1.710000e-02)
    (23, 1.180000e-02)
    (24, 5.660000e-03)
    (25, 2.900000e-03)
    (26, 1.190000e-03)
    (27, 4.460000e-04)
    (28, 1.980000e-04)
    (29, 1.140000e-04)
    (30, 4.380000e-05)
    (31, 1.580000e-05)
    (32, 7.910000e-06)
    (33, 0)
    (34, 0)
    (35, 0)
    (36, 0)
    };

\addplot[
    color=blue,
    mark=triangle*,
    ]
    coordinates {
    (0, 2.930000e-01)
    (1, 2.820000e-01)
    (2, 2.660000e-01)
    (3, 2.620000e-01)
    (4, 2.510000e-01)
    (5, 2.420000e-01)
    (6, 2.320000e-01)
    (7, 2.180000e-01)
    (8, 2.050000e-01)
    (9, 2.000000e-01)
    (10, 1.890000e-01)
    (11, 1.740000e-01)
    (12, 1.660000e-01)
    (13, 1.480000e-01)
    (14, 1.260000e-01)
    (15, 1.120000e-01)
    (16, 8.630000e-02)
    (17, 5.830000e-02)
    (18, 4.240000e-02)
    (19, 2.180000e-02)
    (20, 9.380000e-03)
    (21, 4.450000e-03)
    (22, 2.500000e-03)
    (23, 1.140000e-03)
    (24, 5.110000e-04)
    (25, 2.940000e-04)
    (26, 2.410000e-04)
    (27, 2.340000e-04)
    (28, 1.730000e-04)
    (29, 1.290000e-04)
    (30, 9.450000e-05)
    (31, 6.810000e-05)
    (32, 4.000000e-05)
    (33, 2.580000e-05)
    (34, 2.120000e-05)
    (35, 1.210000e-05)
    (36, 8.450000e-06)
    };
    
\addplot[
    color=green,
    mark=diamond*,
    ]
    coordinates {
    (0, 2.970000e-01)
    (1, 2.800000e-01)
    (2, 2.700000e-01)
    (3, 2.630000e-01)
    (4, 2.470000e-01)
    (5, 2.350000e-01)
    (6, 2.290000e-01)
    (7, 2.300000e-01)
    (8, 2.120000e-01)
    (9, 1.860000e-01)
    (10, 1.880000e-01)
    (11, 1.750000e-01)
    (12, 1.680000e-01)
    (13, 1.360000e-01)
    (14, 1.260000e-01)
    (15, 1.240000e-01)
    (16, 7.830000e-02)
    (17, 6.570000e-02)
    (18, 3.550000e-02)
    (19, 2.180000e-02)
    (20, 1.420000e-02)
    (21, 4.720000e-03)
    (22, 1.900000e-03)
    (23, 1.340000e-03)
    (24, 6.590000e-04)
    (25, 2.750000e-04)
    (26, 1.500000e-04)
    (27, 8.120000e-05)
    (28, 4.160000e-05)
    (29, 3.170000e-05)
    (30, 1.440000e-05)
    (31, 1.300000e-05)
    (32, 1.020000e-05)
    (33, 6.920000e-06)
    (34, 0)
    (35, 0)
    (36, 0)
    };

\addplot[
    color=orange,
    mark=pentagon*,
    ]
    coordinates {
    (0, 2.790000e-01)
    (1, 2.730000e-01)
    (2, 2.590000e-01)
    (3, 2.550000e-01)
    (4, 2.420000e-01)
    (5, 2.310000e-01)
    (6, 2.250000e-01)
    (7, 2.080000e-01)
    (8, 1.980000e-01)
    (9, 1.880000e-01)
    (10, 1.770000e-01)
    (11, 1.550000e-01)
    (12, 1.500000e-01)
    (13, 1.290000e-01)
    (14, 1.060000e-01)
    (15, 8.090000e-02)
    (16, 5.860000e-02)
    (17, 4.120000e-02)
    (18, 1.890000e-02)
    (19, 8.380000e-03)
    (20, 4.310000e-03)
    (21, 2.470000e-03)
    (22, 9.240000e-04)
    (23, 3.850000e-04)
    (24, 2.420000e-04)
    (25, 1.220000e-04)
    (26, 7.330000e-05)
    (27, 4.260000e-05)
    (28, 2.570000e-05)
    (29, 1.660000e-05)
    (30, 1.120000e-05)
    (31, 7.300000e-06)
    (32, 0)
    (33, 0)
    (34, 0)
    (35, 0)
    (36, 0)
    };
    
\end{axis}
\end{tikzpicture}
\captionsetup{justification=centering}
\caption{\small MCS2 (QPSK, CR = 3/4)}
\end{subfigure}
\hfill
\begin{subfigure}[h]{0.32\textwidth}
\vspace{-0.4cm}
\begin{tikzpicture}[font=\small]
\begin{axis}[
    width=1\linewidth,
    height=0.92\linewidth,
    ymode=log,
    xlabel={SNR [dB]},
    ylabel={BER},
    xmin=5, xmax=35,
    ymin=1e-5, ymax=1,
    xtick={0,5, 10, 15, 20, 25, 30, 35},
    legend pos=south west,
    xmajorgrids=true,
    ymajorgrids=true,
    grid style=dashed,
]

\addplot[
    color=red,
    mark=square*,
    ]
    coordinates {
(0, 3.600000e-01)
(1, 3.550000e-01)
(2, 3.490000e-01)
(3, 3.340000e-01)
(4, 3.300000e-01)
(5, 3.240000e-01)
(6, 3.190000e-01)
(7, 3.150000e-01)
(8, 3.000000e-01)
(9, 2.980000e-01)
(10, 2.910000e-01)
(11, 2.800000e-01)
(12, 2.680000e-01)
(13, 2.640000e-01)
(14, 2.450000e-01)
(15, 2.380000e-01)
(16, 2.230000e-01)
(17, 2.040000e-01)
(18, 1.920000e-01)
(19, 1.530000e-01)
(20, 1.250000e-01)
(21, 9.720000e-02)
(22, 6.500000e-02)
(23, 3.670000e-02)
(24, 2.210000e-02)
(25, 1.130000e-02)
(26, 4.840000e-03)
(27, 2.520000e-03)
(28, 9.770000e-04)
(29, 5.420000e-04)
(30, 1.920000e-04)
(31, 7.250000e-05)
(32, 3.930000e-05)
(33, 1.590000e-05)
(34, 5.740000e-06)
    };

\addplot[
    color=blue,
    mark=triangle*,
    ]
    coordinates {
(0, 3.620000e-01)
(1, 3.520000e-01)
(2, 3.520000e-01)
(3, 3.350000e-01)
(4, 3.320000e-01)
(5, 3.270000e-01)
(6, 3.230000e-01)
(7, 3.130000e-01)
(8, 2.990000e-01)
(9, 2.980000e-01)
(10, 2.780000e-01)
(11, 2.710000e-01)
(12, 2.560000e-01)
(13, 2.430000e-01)
(14, 2.260000e-01)
(15, 2.140000e-01)
(16, 1.730000e-01)
(17, 1.440000e-01)
(18, 8.240000e-02)
(19, 5.070000e-02)
(20, 2.140000e-02)
(21, 8.040000e-03)
(22, 2.890000e-03)
(23, 7.340000e-04)
(24, 3.720000e-04)
(25, 8.540000e-05)
(26, 3.560000e-05)
(27, 9.510000e-06)
(28, 0)
(29, 0)
(30, 0)
(31, 0)
(32, 0)
(33, 0)
(34, 0)
    };
    
\addplot[
    color=green,
    mark=diamond*,
    ]
    coordinates {
(0, 3.600000e-01)
(1, 3.540000e-01)
(2, 3.520000e-01)
(3, 3.380000e-01)
(4, 3.300000e-01)
(5, 3.260000e-01)
(6, 3.220000e-01)
(7, 3.150000e-01)
(8, 3.040000e-01)
(9, 2.960000e-01)
(10, 2.870000e-01)
(11, 2.700000e-01)
(12, 2.600000e-01)
(13, 2.480000e-01)
(14, 2.320000e-01)
(15, 2.190000e-01)
(16, 1.820000e-01)
(17, 1.470000e-01)
(18, 9.620000e-02)
(19, 6.540000e-02)
(20, 2.580000e-02)
(21, 1.110000e-02)
(22, 3.500000e-03)
(23, 9.000000e-04)
(24, 3.570000e-04)
(25, 8.400000e-05)
(26, 2.580000e-05)
(27, 6.830000e-06)
(28, 0)
(29, 0)
(30, 0)
(31, 0)
(32, 0)
(33, 0)
(34, 0)
    };

\addplot[
    color=orange,
    mark=pentagon*,
    ]
    coordinates {
(0, 3.630000e-01)
(1, 3.560000e-01)
(2, 3.530000e-01)
(3, 3.450000e-01)
(4, 3.340000e-01)
(5, 3.270000e-01)
(6, 3.160000e-01)
(7, 3.160000e-01)
(8, 3.000000e-01)
(9, 2.970000e-01)
(10, 2.890000e-01)
(11, 2.760000e-01)
(12, 2.590000e-01)
(13, 2.570000e-01)
(14, 2.390000e-01)
(15, 2.200000e-01)
(16, 1.900000e-01)
(17, 1.540000e-01)
(18, 1.140000e-01)
(19, 6.470000e-02)
(20, 3.270000e-02)
(21, 1.300000e-02)
(22, 4.420000e-03)
(23, 1.150000e-03)
(24, 4.140000e-04)
(25, 1.100000e-04)
(26, 2.790000e-05)
(27, 7.850000e-06)
(28, 0)
(29, 0)
(30, 0)
(31, 0)
(32, 0)
(33, 0)
(34, 0)
    };
    
\end{axis}
\end{tikzpicture}
\captionsetup{justification=centering}
\caption{\small MCS3 (8-PSK, CR = 1/2)}
\end{subfigure}
\hfill
\begin{subfigure}[h]{0.32\textwidth}
\begin{tikzpicture}[font=\small]
\begin{axis}[
    width=1\linewidth,
    height=0.92\linewidth,
    legend style={fill opacity=0.5, text opacity=1, draw opacity=1},
    ymode=log,
    xlabel={SNR [dB]},
    ylabel={BER},
    xmin=10, xmax=50,
    ymin=1e-5, ymax=1,
    xtick={0,5, 10, 15, 20, 25, 30, 35, 40, 45, 50},
    legend pos=south west,
    xmajorgrids=true,
    ymajorgrids=true,
    grid style=dashed,
]

\addplot[
    color=red,
    mark=square*,
    ]
    coordinates {
(0, 3.490000e-01)
(1, 3.390000e-01)
(2, 3.290000e-01)
(3, 3.250000e-01)
(4, 3.170000e-01)
(5, 3.070000e-01)
(6, 3.000000e-01)
(7, 2.950000e-01)
(8, 2.860000e-01)
(9, 2.760000e-01)
(10, 2.700000e-01)
(11, 2.580000e-01)
(12, 2.510000e-01)
(13, 2.450000e-01)
(14, 2.310000e-01)
(15, 2.200000e-01)
(16, 2.130000e-01)
(17, 2.080000e-01)
(18, 1.960000e-01)
(19, 1.920000e-01)
(20, 1.850000e-01)
(21, 1.710000e-01)
(22, 1.640000e-01)
(23, 1.570000e-01)
(24, 1.460000e-01)
(25, 1.400000e-01)
(26, 1.290000e-01)
(27, 1.170000e-01)
(28, 1.080000e-01)
(29, 8.400000e-02)
(30, 7.610000e-02)
(31, 6.440000e-02)
(32, 4.670000e-02)
(33, 3.690000e-02)
(34, 2.700000e-02)
(35, 2.160000e-02)
(36, 1.400000e-02)
(37, 8.800000e-03)
(38, 6.060000e-03)
(39, 3.810000e-03)
(40, 2.750000e-03)
(41, 1.520000e-03)
(42, 9.790000e-04)
(43, 5.350000e-04)
(44, 4.430000e-04)
(45, 2.570000e-04)
(46, 1.650000e-04)
(47, 9.270000e-05)
(48, 5.300000e-05)
(49, 2.920000e-05)
(50, 2.200000e-05)
    };

\addplot[
    color=blue,
    mark=triangle*,
    ]
    coordinates {
(0, 3.520000e-01)
(1, 3.360000e-01)
(2, 3.280000e-01)
(3, 3.250000e-01)
(4, 3.170000e-01)
(5, 3.040000e-01)
(6, 2.990000e-01)
(7, 2.940000e-01)
(8, 2.800000e-01)
(9, 2.730000e-01)
(10, 2.610000e-01)
(11, 2.480000e-01)
(12, 2.430000e-01)
(13, 2.300000e-01)
(14, 2.140000e-01)
(15, 2.020000e-01)
(16, 1.940000e-01)
(17, 1.760000e-01)
(18, 1.640000e-01)
(19, 1.530000e-01)
(20, 1.410000e-01)
(21, 1.240000e-01)
(22, 1.040000e-01)
(23, 7.290000e-02)
(24, 5.850000e-02)
(25, 3.450000e-02)
(26, 1.670000e-02)
(27, 8.410000e-03)
(28, 3.010000e-03)
(29, 1.690000e-03)
(30, 6.380000e-04)
(31, 3.410000e-04)
(32, 1.760000e-04)
(33, 7.610000e-05)
(34, 3.620000e-05)
(35, 1.590000e-05)
(36, 1.120000e-05)
(37, 7.290000e-06)
(38, 0)
(39, 0)
(40, 0)
(41, 0)
(42, 0)
(43, 0)
(44, 0)
(45, 0)
(46, 0)
(47, 0)
(48, 0)
(49, 0)
(50, 0)
    };
    
\addplot[
    color=green,
    mark=diamond*,
    ]
    coordinates {
(0, 3.480000e-01)
(1, 3.450000e-01)
(2, 3.360000e-01)
(3, 3.240000e-01)
(4, 3.100000e-01)
(5, 3.110000e-01)
(6, 3.060000e-01)
(7, 2.900000e-01)
(8, 2.850000e-01)
(9, 2.750000e-01)
(10, 2.640000e-01)
(11, 2.550000e-01)
(12, 2.480000e-01)
(13, 2.290000e-01)
(14, 2.290000e-01)
(15, 2.040000e-01)
(16, 2.020000e-01)
(17, 1.900000e-01)
(18, 1.680000e-01)
(19, 1.580000e-01)
(20, 1.490000e-01)
(21, 1.290000e-01)
(22, 9.780000e-02)
(23, 8.800000e-02)
(24, 5.720000e-02)
(25, 3.090000e-02)
(26, 2.010000e-02)
(27, 8.500000e-03)
(28, 4.440000e-03)
(29, 1.870000e-03)
(30, 5.540000e-04)
(31, 2.820000e-04)
(32, 1.550000e-04)
(33, 8.450000e-05)
(34, 2.860000e-05)
(35, 1.540000e-05)
(36, 7.400000e-06)
(37, 0)
(38, 0)
(39, 0)
(40, 0)
(41, 0)
(42, 0)
(43, 0)
(44, 0)
(45, 0)
(46, 0)
(47, 0)
(48, 0)
(49, 0)
(50, 0)
    };

\addplot[
    color=orange,
    mark=pentagon*,
    ]
    coordinates {
(0, 3.440000e-01)
(1, 3.340000e-01)
(2, 3.230000e-01)
(3, 3.170000e-01)
(4, 3.130000e-01)
(5, 2.990000e-01)
(6, 2.940000e-01)
(7, 2.880000e-01)
(8, 2.750000e-01)
(9, 2.710000e-01)
(10, 2.570000e-01)
(11, 2.460000e-01)
(12, 2.360000e-01)
(13, 2.260000e-01)
(14, 2.140000e-01)
(15, 1.950000e-01)
(16, 1.850000e-01)
(17, 1.740000e-01)
(18, 1.580000e-01)
(19, 1.440000e-01)
(20, 1.300000e-01)
(21, 1.100000e-01)
(22, 9.370000e-02)
(23, 6.000000e-02)
(24, 4.050000e-02)
(25, 1.940000e-02)
(26, 9.520000e-03)
(27, 3.850000e-03)
(28, 1.530000e-03)
(29, 6.790000e-04)
(30, 2.960000e-04)
(31, 1.710000e-04)
(32, 7.490000e-05)
(33, 3.680000e-05)
(34, 1.500000e-05)
(35, 9.190000e-06)
(36, 0)
(37, 0)
(38, 0)
(39, 0)
(40, 0)
(41, 0)
(42, 0)
(43, 0)
(44, 0)
(45, 0)
(46, 0)
(47, 0)
(48, 0)
(49, 0)
(50, 0)
    };
    
\end{axis}
\end{tikzpicture}
\captionsetup{justification=centering}
\caption{\small MCS4 (8-PSK, CR = 3/4)}
\end{subfigure}
\hfill
\begin{subfigure}[h]{0.32\textwidth}
\begin{tikzpicture}[font=\small]
\begin{axis}[
    width=1\linewidth,
    height=0.92\linewidth,
    ymode=log,
    xlabel={SNR [dB]},
    ylabel={BER},
    xmin=10, xmax=50,
    ymin=1e-5, ymax=1,
    xtick={0,5, 10, 15, 20, 25, 30, 35, 40, 45, 50},
    legend pos=south west,
    xmajorgrids=true,
    ymajorgrids=true,
    grid style=dashed,
]

\addplot[
    color=red,
    mark=square*,
    ]
    coordinates {
(0, 3.890000e-01)
(1, 3.760000e-01)
(2, 3.720000e-01)
(3, 3.720000e-01)
(4, 3.600000e-01)
(5, 3.530000e-01)
(6, 3.510000e-01)
(7, 3.470000e-01)
(8, 3.400000e-01)
(9, 3.300000e-01)
(10, 3.290000e-01)
(11, 3.170000e-01)
(12, 3.110000e-01)
(13, 3.100000e-01)
(14, 2.960000e-01)
(15, 2.920000e-01)
(16, 2.820000e-01)
(17, 2.770000e-01)
(18, 2.630000e-01)
(19, 2.510000e-01)
(20, 2.440000e-01)
(21, 2.400000e-01)
(22, 2.180000e-01)
(23, 2.030000e-01)
(24, 1.700000e-01)
(25, 1.490000e-01)
(26, 1.140000e-01)
(27, 8.390000e-02)
(28, 6.840000e-02)
(29, 4.150000e-02)
(30, 2.650000e-02)
(31, 1.580000e-02)
(32, 1.040000e-02)
(33, 6.510000e-03)
(34, 3.170000e-03)
(35, 2.070000e-03)
(36, 7.700000e-04)
(37, 6.060000e-04)
(38, 2.530000e-04)
(39, 1.390000e-04)
(40, 7.560000e-05)
(41, 4.740000e-05)
(42, 2.360000e-05)
(43, 1.200000e-05)
(44, 5.760000e-06)
(45, 0)
(46, 0)
(47, 0)
(48, 0)
(49, 0)
(50, 0)
    };

\addplot[
    color=blue,
    mark=triangle*,
    ]
    coordinates {
(0, 3.840000e-01)
(1, 3.770000e-01)
(2, 3.700000e-01)
(3, 3.740000e-01)
(4, 3.670000e-01)
(5, 3.530000e-01)
(6, 3.520000e-01)
(7, 3.420000e-01)
(8, 3.360000e-01)
(9, 3.280000e-01)
(10, 3.250000e-01)
(11, 3.140000e-01)
(12, 3.090000e-01)
(13, 3.030000e-01)
(14, 2.960000e-01)
(15, 2.800000e-01)
(16, 2.680000e-01)
(17, 2.640000e-01)
(18, 2.460000e-01)
(19, 2.330000e-01)
(20, 2.240000e-01)
(21, 1.980000e-01)
(22, 1.580000e-01)
(23, 1.280000e-01)
(24, 8.310000e-02)
(25, 4.250000e-02)
(26, 3.050000e-02)
(27, 1.520000e-02)
(28, 6.240000e-03)
(29, 3.450000e-03)
(30, 1.280000e-03)
(31, 5.440000e-04)
(32, 2.160000e-04)
(33, 1.020000e-04)
(34, 8.070000e-05)
(35, 5.510000e-05)
(36, 6.690000e-05)
(37, 1.460000e-04)
(38, 3.620000e-04)
(39, 5.800000e-04)
(40, 1.150000e-03)
(41, 1.850000e-03)
(42, 2.400000e-03)
(43, 3.080000e-03)
(44, 3.720000e-03)
(45, 3.810000e-03)
(46, 5.340000e-03)
(47, 4.820000e-03)
(48, 5.360000e-03)
(49, 6.000000e-03)
(50, 4.350000e-03)
    };
    
\addplot[
    color=green,
    mark=diamond*,
    ]
    coordinates {
(0, 3.970000e-01)
(1, 3.880000e-01)
(2, 3.860000e-01)
(3, 3.780000e-01)
(4, 3.800000e-01)
(5, 3.660000e-01)
(6, 3.610000e-01)
(7, 3.520000e-01)
(8, 3.420000e-01)
(9, 3.390000e-01)
(10, 3.310000e-01)
(11, 3.200000e-01)
(12, 3.180000e-01)
(13, 3.130000e-01)
(14, 2.990000e-01)
(15, 2.880000e-01)
(16, 2.780000e-01)
(17, 2.740000e-01)
(18, 2.560000e-01)
(19, 2.410000e-01)
(20, 2.330000e-01)
(21, 2.140000e-01)
(22, 1.760000e-01)
(23, 1.500000e-01)
(24, 1.060000e-01)
(25, 6.100000e-02)
(26, 4.420000e-02)
(27, 2.060000e-02)
(28, 1.360000e-02)
(29, 5.790000e-03)
(30, 4.920000e-03)
(31, 1.930000e-03)
(32, 8.030000e-04)
(33, 4.560000e-04)
(34, 2.150000e-04)
(35, 1.400000e-04)
(36, 6.800000e-05)
(37, 5.290000e-05)
(38, 2.460000e-05)
(39, 1.560000e-05)
(40, 1.410000e-05)
(41, 9.260000e-06)
(42, 0)
(43, 0)
(44, 0)
(45, 0)
(46, 0)
(47, 0)
(48, 0)
(49, 0)
(50, 0)
    };

\addplot[
    color=orange,
    mark=pentagon*,
    ]
    coordinates {
(0, 3.970000e-01)
(1, 3.860000e-01)
(2, 3.830000e-01)
(3, 3.800000e-01)
(4, 3.730000e-01)
(5, 3.570000e-01)
(6, 3.550000e-01)
(7, 3.530000e-01)
(8, 3.460000e-01)
(9, 3.370000e-01)
(10, 3.370000e-01)
(11, 3.210000e-01)
(12, 3.140000e-01)
(13, 3.090000e-01)
(14, 3.020000e-01)
(15, 2.900000e-01)
(16, 2.780000e-01)
(17, 2.780000e-01)
(18, 2.570000e-01)
(19, 2.440000e-01)
(20, 2.310000e-01)
(21, 2.220000e-01)
(22, 1.720000e-01)
(23, 1.470000e-01)
(24, 1.140000e-01)
(25, 6.260000e-02)
(26, 4.090000e-02)
(27, 1.990000e-02)
(28, 1.270000e-02)
(29, 6.250000e-03)
(30, 4.600000e-03)
(31, 1.780000e-03)
(32, 8.500000e-04)
(33, 4.130000e-04)
(34, 1.610000e-04)
(35, 1.150000e-04)
(36, 6.520000e-05)
(37, 3.800000e-05)
(38, 1.540000e-05)
(39, 1.400000e-05)
(40, 8.140000e-06)
(41, 0)
(42, 0)
(43, 0)
(44, 0)
(45, 0)
(46, 0)
(47, 0)
(48, 0)
(49, 0)
(50, 0)
    };
    
\end{axis}
\end{tikzpicture}
\captionsetup{justification=centering}
\caption{\small MCS5 (16-QAM, CR = 1/2)}
\end{subfigure}
\hfill
\begin{subfigure}[h]{0.32\textwidth}
\begin{tikzpicture}[font=\small]
\begin{axis}[
    width=1\linewidth,
    height=0.92\linewidth,
    ymode=log,
    xlabel={SNR [dB]},
    ylabel={BER},
    xmin=10, xmax=50, 
    ymin=1e-5, ymax=1,
    xtick={0, 5, 10, 15, 20, 25, 30, 35, 40, 45, 50},
    legend pos=south west,
    xmajorgrids=true,
    ymajorgrids=true,
    grid style=dashed,
]

\addplot[
    color=red,
    mark=square*,
    ]
    coordinates {
(0, 3.770000e-01)
(1, 3.680000e-01)
(2, 3.670000e-01)
(3, 3.590000e-01)
(4, 3.480000e-01)
(5, 3.480000e-01)
(6, 3.380000e-01)
(7, 3.270000e-01)
(8, 3.220000e-01)
(9, 3.170000e-01)
(10, 3.090000e-01)
(11, 2.990000e-01)
(12, 2.910000e-01)
(13, 2.820000e-01)
(14, 2.720000e-01)
(15, 2.650000e-01)
(16, 2.570000e-01)
(17, 2.510000e-01)
(18, 2.440000e-01)
(19, 2.330000e-01)
(20, 2.240000e-01)
(21, 2.160000e-01)
(22, 2.080000e-01)
(23, 1.980000e-01)
(24, 1.940000e-01)
(25, 1.840000e-01)
(26, 1.780000e-01)
(27, 1.650000e-01)
(28, 1.580000e-01)
(29, 1.530000e-01)
(30, 1.470000e-01)
(31, 1.390000e-01)
(32, 1.280000e-01)
(33, 1.170000e-01)
(34, 1.090000e-01)
(35, 9.600000e-02)
(36, 8.390000e-02)
(37, 8.270000e-02)
(38, 5.120000e-02)
(39, 4.760000e-02)
(40, 4.230000e-02)
(41, 2.890000e-02)
(42, 2.400000e-02)
(43, 1.940000e-02)
(44, 1.330000e-02)
(45, 1.150000e-02)
(46, 6.500000e-03)
(47, 5.080000e-03)
(48, 3.670000e-03)
(49, 2.520000e-03)
(50, 1.600000e-03)
(51, 1.190000e-03)
(52, 9.050000e-04)
(53, 8.540000e-04)
(54, 5.580000e-04)
(55, 4.250000e-04)
(56, 2.870000e-04)
(57, 2.220000e-04)
(58, 1.520000e-04)
(59, 1.180000e-04)
(60, 8.410000e-05)
(61, 6.520000e-05)
(62, 4.070000e-05)
(63, 3.070000e-05)
(64, 2.010000e-05)
(65, 1.220000e-05)
(66, 6.920000e-06)
(67, 0)
(68, 0)
(69, 0)
(70, 0)
(71, 0)
(72, 0)
(73, 0)
(74, 0)
(75, 0)
(76, 0)
(77, 0)
(78, 0)
(79, 0)
    };

\addplot[
    color=blue,
    mark=triangle*,
    ]
    coordinates {
(0, 3.760000e-01)
(1, 3.700000e-01)
(2, 3.680000e-01)
(3, 3.590000e-01)
(4, 3.470000e-01)
(5, 3.450000e-01)
(6, 3.380000e-01)
(7, 3.250000e-01)
(8, 3.170000e-01)
(9, 3.140000e-01)
(10, 3.050000e-01)
(11, 2.970000e-01)
(12, 2.880000e-01)
(13, 2.800000e-01)
(14, 2.650000e-01)
(15, 2.570000e-01)
(16, 2.480000e-01)
(17, 2.400000e-01)
(18, 2.330000e-01)
(19, 2.200000e-01)
(20, 2.060000e-01)
(21, 2.000000e-01)
(22, 1.890000e-01)
(23, 1.770000e-01)
(24, 1.680000e-01)
(25, 1.550000e-01)
(26, 1.460000e-01)
(27, 1.260000e-01)
(28, 1.190000e-01)
(29, 1.050000e-01)
(30, 7.730000e-02)
(31, 6.570000e-02)
(32, 5.050000e-02)
(33, 3.570000e-02)
(34, 2.940000e-02)
(35, 2.390000e-02)
(36, 1.810000e-02)
(37, 1.640000e-02)
(38, 1.370000e-02)
(39, 1.360000e-02)
(40, 1.510000e-02)
(41, 1.220000e-02)
(42, 1.280000e-02)
(43, 1.210000e-02)
(44, 1.510000e-02)
(45, 1.190000e-02)
(46, 1.180000e-02)
(47, 1.020000e-02)
(48, 1.030000e-02)
(49, 1.190000e-02)
(50, 1.100000e-02)
(51, 1.200000e-02)
(52, 1.220000e-02)
(53, 9.750000e-03)
(54, 1.080000e-02)
(55, 1.290000e-02)
(56, 1.040000e-02)
(57, 1.040000e-02)
(58, 1.210000e-02)
(59, 1.210000e-02)
(60, 1.050000e-02)
(61, 1.070000e-02)
(62, 8.850000e-03)
(63, 8.350000e-03)
(64, 7.460000e-03)
(65, 5.400000e-03)
(66, 3.670000e-03)
(67, 3.550000e-03)
(68, 2.940000e-03)
(69, 2.240000e-03)
(70, 1.750000e-03)
(71, 1.130000e-03)
(72, 6.910000e-04)
(73, 3.610000e-04)
(74, 1.940000e-04)
(75, 1.460000e-04)
(76, 1.160000e-04)
(77, 7.670000e-05)
(78, 2.450000e-05)
(79, 2.700000e-06)
    };
    
\addplot[
    color=green,
    mark=diamond*,
    ]
    coordinates {
(0, 3.910000e-01)
(1, 3.850000e-01)
(2, 3.820000e-01)
(3, 3.710000e-01)
(4, 3.620000e-01)
(5, 3.550000e-01)
(6, 3.490000e-01)
(7, 3.350000e-01)
(8, 3.290000e-01)
(9, 3.240000e-01)
(10, 3.130000e-01)
(11, 3.050000e-01)
(12, 2.960000e-01)
(13, 2.860000e-01)
(14, 2.700000e-01)
(15, 2.640000e-01)
(16, 2.560000e-01)
(17, 2.450000e-01)
(18, 2.380000e-01)
(19, 2.220000e-01)
(20, 2.110000e-01)
(21, 2.030000e-01)
(22, 1.910000e-01)
(23, 1.800000e-01)
(24, 1.710000e-01)
(25, 1.640000e-01)
(26, 1.540000e-01)
(27, 1.340000e-01)
(28, 1.260000e-01)
(29, 1.160000e-01)
(30, 1.020000e-01)
(31, 7.900000e-02)
(32, 6.760000e-02)
(33, 5.100000e-02)
(34, 4.970000e-02)
(35, 3.270000e-02)
(36, 3.040000e-02)
(37, 2.430000e-02)
(38, 1.830000e-02)
(39, 1.570000e-02)
(40, 9.640000e-03)
(41, 9.480000e-03)
(42, 6.420000e-03)
(43, 5.040000e-03)
(44, 4.480000e-03)
(45, 4.250000e-03)
(46, 2.290000e-03)
(47, 1.870000e-03)
(48, 1.450000e-03)
(49, 1.310000e-03)
(50, 1.250000e-03)
(51, 9.430000e-04)
(52, 8.050000e-04)
(53, 6.390000e-04)
(54, 4.600000e-04)
(55, 4.130000e-04)
(56, 3.390000e-04)
(57, 2.910000e-04)
(58, 2.230000e-04)
(59, 1.840000e-04)
(60, 1.700000e-04)
(61, 1.010000e-04)
(62, 9.540000e-05)
(63, 6.320000e-05)
(64, 4.830000e-05)
(65, 2.670000e-05)
(66, 1.840000e-05)
(67, 1.210000e-05)
(68, 6.660000e-06)
(69, 0)
(70, 0)
(71, 0)
(72, 0)
(73, 0)
(74, 0)
(75, 0)
(76, 0)
(77, 0)
(78, 0)
(79, 0)
    };

\addplot[
    color=orange,
    mark=pentagon*,
    ]
    coordinates {
(0, 3.920000e-01)
(1, 3.870000e-01)
(2, 3.800000e-01)
(3, 3.720000e-01)
(4, 3.630000e-01)
(5, 3.570000e-01)
(6, 3.500000e-01)
(7, 3.370000e-01)
(8, 3.290000e-01)
(9, 3.240000e-01)
(10, 3.150000e-01)
(11, 3.050000e-01)
(12, 2.940000e-01)
(13, 2.870000e-01)
(14, 2.720000e-01)
(15, 2.630000e-01)
(16, 2.550000e-01)
(17, 2.450000e-01)
(18, 2.390000e-01)
(19, 2.210000e-01)
(20, 2.130000e-01)
(21, 2.040000e-01)
(22, 1.950000e-01)
(23, 1.800000e-01)
(24, 1.740000e-01)
(25, 1.600000e-01)
(26, 1.500000e-01)
(27, 1.330000e-01)
(28, 1.270000e-01)
(29, 1.180000e-01)
(30, 9.730000e-02)
(31, 8.270000e-02)
(32, 6.720000e-02)
(33, 5.490000e-02)
(34, 4.880000e-02)
(35, 3.570000e-02)
(36, 2.850000e-02)
(37, 2.430000e-02)
(38, 2.040000e-02)
(39, 1.570000e-02)
(40, 1.030000e-02)
(41, 9.620000e-03)
(42, 6.980000e-03)
(43, 5.450000e-03)
(44, 4.660000e-03)
(45, 3.840000e-03)
(46, 2.510000e-03)
(47, 1.870000e-03)
(48, 1.640000e-03)
(49, 1.390000e-03)
(50, 1.350000e-03)
(51, 9.080000e-04)
(52, 8.450000e-04)
(53, 6.140000e-04)
(54, 4.680000e-04)
(55, 4.640000e-04)
(56, 3.140000e-04)
(57, 2.940000e-04)
(58, 2.070000e-04)
(59, 1.720000e-04)
(60, 1.470000e-04)
(61, 8.960000e-05)
(62, 8.160000e-05)
(63, 6.860000e-05)
(64, 4.550000e-05)
(65, 2.410000e-05)
(66, 1.890000e-05)
(67, 1.060000e-05)
(68, 6.190000e-06)
(69, 0)
(70, 0)
(71, 0)
(72, 0)
(73, 0)
(74, 0)
(75, 0)
(76, 0)
(77, 0)
(78, 0)
(79, 0)
    };
    \legend{Original S=0, Original S=1, Simplified S=1, Fixed-Point S=1}
    
\end{axis}
\end{tikzpicture}
\captionsetup{justification=centering}
\caption{\small MCS6 (16-QAM, CR = 3/4)}
\end{subfigure}
\caption{Impact in terms of performance of simplification and fixed-point conversion.}
\label{fig:fixed_point}
\end{figure*}

Fixed-point conversion of the soft mapping/demapping algorithms was done using Fxpmath \cite{FXP_math20}, a Python library for fractional fixed-point (base 2) arithmetic and binary manipulation with Numpy \cite{numpy20} compatibility. The focus of the conversion was on the calculations for bitwise max-log MAP demapper, bitwise soft mapper, and EP-based soft estimates (see Fig.~\ref{fig_bitwiseDemap}), to facilitate the soft mapper/demapper architecture design.

For each variable of the different functions, we determined whether the variable is signed ($Q_S=0$ or $1$), the number of integer bits $Q_I$, and the number of fractional bits $Q_F$. The number of integer bits was determined by analyzing the maximum absolute value of the variable while operating in different Signal-to-Noise Ratios (SNRs) and scenarios. Then, an empirical approach was applied thanks to the py-AFF3CT environment. Indeed, this toolbox enables to rapidly estimate if reducing the number of bits would impact the Bit Error Rate (BER). A similar process was carried out for the fractional bits, comparing the impact on BER of different sizes of fractional parts to achieve the minimum size with negligible loss. The total bit size of a variable is given by $Q_T = Q_S + Q_I + Q_F$.

As an example, consider the equalized symbol estimates $x_k^{e(s)}$, which are used in the simplified soft demapping (Sect.~\ref{simp_soft_demap}). For the case of the Proakis-C channel with AWGN for QPSK, this variable has a maximum value of around 6.3 at 0dB, 3 at 10dB, and 2.2 at 20dB. 
Therefore, the performances were compared for the cases of having many bits for the fractional part $Q_F$ (16 bits) while having 3, 2, 1, and 0 bits for the integer part $Q_I$.

The tests showed that with $Q_I = 0$, a loss of around 0.5dB at BER $10^{-3}$ is introduced. For the other values, there is negligible loss when using 1 integer bit compared to 2 and 3. 
This is due to noise amplitude having a more important range than the normalized signals at low SNRs. 
As for the fractional part, for QPSK, there is not a clear loss introduced with 3 bits. However, for 8-PSK, the use of only 3 fractional bits introduces a loss of around 1dB at BER $10^{-3}$. With $Q_F = 4$, there is a smaller but still perceptible loss of around 0.1dB.

Considering these empirical observations, 2 bits were allocated for the integer part and 5 bits for the fractional part, as a safe margin for other channels and scenarios, and 1 bit for the sign. In the end, $Q_T=8$ bits were assigned to the representation of the symbol estimates. 
The sizes of all the other variables were chosen with a similar process. All the named variables were successfully converted to $Q_T = 8$, while some internal calculations need up to 10 bits.

The performances in terms of Bit Error Rate (BER) for the original algorithm, the simplified algorithm, and its fixed-point version are shown in Fig.~\ref{fig:fixed_point}. 
The examined channel was the challenging Proakis-C. We focused on six different Modulation and Coding Schemes (MCS): the three presented constellations, using the LTE turbo code and rate matching with code rates $1/2$ and $3/4$. All the simulations were done considering one self-iteration ($S=1$). The standard linear MMSE equalizer performance (given for $S=0$) is provided for reference.
For all cases, the fixed-point version performs very similarly to the simplified floating-point one. 
For Proakis-C MCS 2, 5, and 6, the original algorithm performs significantly worse than the simplified version at high SNR. 
In fact, 
bad interference patterns can make some observations close to wrong constellation points. In such situations, the original EP receiver generates an over-optimistic EP variance, and the EP feedback will be considered as correct while it is not. 
On the other hand, the simplified receiver calculates the EP variance independently of (possibly wrong) observations, through the $C_{EP}$ LUT. This generates a more conservative feedback, thus avoiding the BER degradations.

As for the small gain of the fixed-point version, this observed behavior can be attributed to the inherent quantization noise introduced by the fixed-point representation. Interestingly, this noise can induce perturbations that effectively reduce the occurrence of unfavorable interference patterns, thereby enhancing the robustness of the algorithm in challenging scenarios.

\section{FPGA Implementation and Prototyping}
\label{Res}

\begin{figure}[!b]
\centerline{\includegraphics[width=0.49\textwidth]{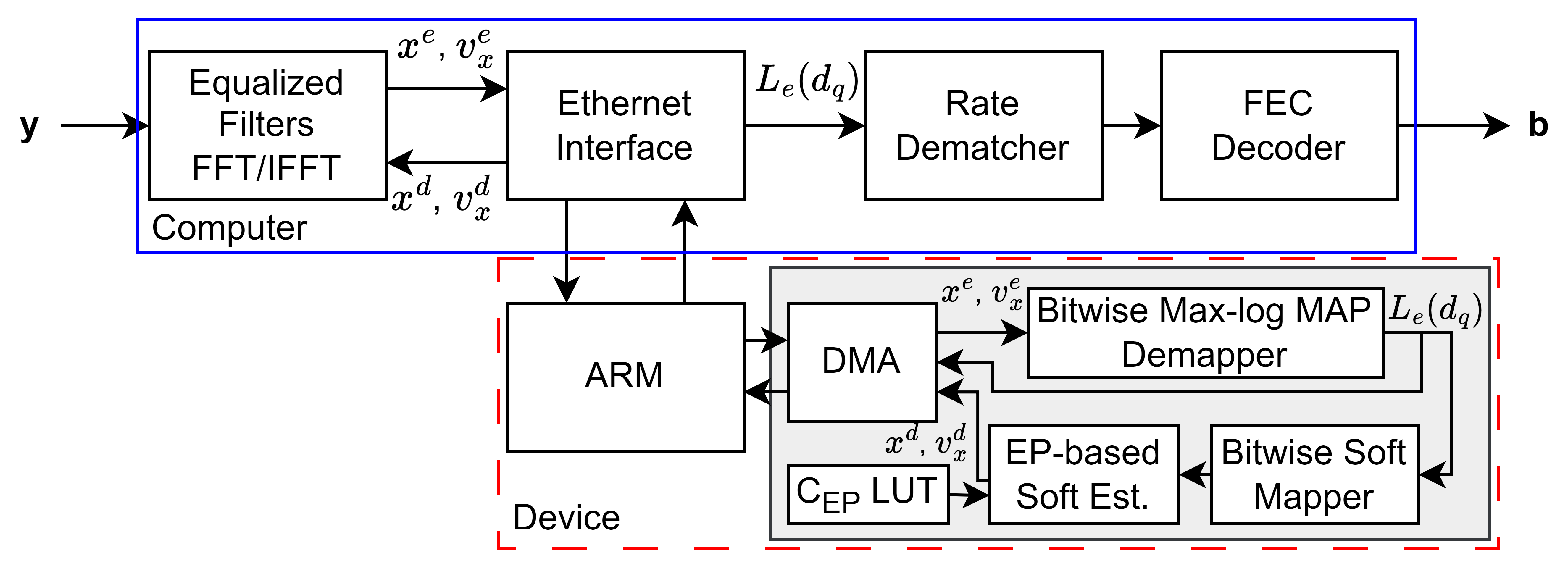}}
\caption{Simplified EP-based FD-SILE FPGA and HIL Block Diagram.}
\label{hil_arch}
\end{figure}

\subsection{Experimental Setup}

The experimental setup employs a Hardware in the Loop (HIL) configuration, involving a computer and a PYNQ Z2 board \cite{pynqz2}. The computer runs py-AFF3CT, a Python wrapper for the Forward Error Correction Toolbox AFF3CT. The PYNQ Z2 board is a prototyping board based on the Xilinx Zynq System on Chip (SoC) with an ARM processor and an FPGA ZYNQ XC7Z020-1CLG400C. In our experimental setup, the board is connected to the computer via an Ethernet cable. The board's ARM processor forwards the data to the FPGA device via a Direct Memory Access (DMA). Moreover, it facilitates the Ethernet connection between the computer and the SoC device, acting as a passthrough.

Fig. \ref{hil_arch} shows the block diagram of simplified EP-based FD-SILE with the HIL configuration. Within the FPGA, the three parts of the algorithm with analytical simplifications have been implemented: bitwise max-log MAP demapper (eq. (\ref{llr_8psk}) and Table \ref{tab_softDemap}), bitwise soft mapper (Table \ref{tab_softMap}) and LUT-based variance parameter ($C_{EP}$ in section \ref{simp_feed_comp}), and EP-based soft estimates (eq. (\ref{var_cep}-\ref{mean_cep})). 
These implemented blocks of the algorithm perform essential computation for the Expectation Propagation-based receiver and are the focus of the presented work. The remaining parts of the receiver are executed thanks to AFF3CT on the associated computer. These software blocks encompass filtering, equalization, FFT and IFFT operations, rate dematching, and Forward Error Correction (FEC) decoding. 
This HIL setup is used to obtain the previously discussed fixed-point simulation results.



\subsection{FPGA Implementation Results}
\label{imp_res}

Table \ref{tab_resources} shows the total available and allocated FPGA resources for each of the three constellations that are considered in this article: QPSK, 8-PSK, and 16-QAM. The first row for each constellation is for the implementation of only FD-SILE without the HIL configuration, while the second row is for the FD-SILE with the HIL configuration, including the data exchanges to the ZYNQ Processing System and the DMA. The implemented architectures use a data width of 32 bits for the DMA connection, enabling to exchange two symbols with 8-bit real and imaginary parts.

Given the low resource usage for the analytical blocks, several calculations can be carried out in parallel in the proposed architectures. For QPSK and 16-QAM, the architectures are composed of four independent bitwise max-log MAP demapper blocks, as the real and imaginary parts are independent, 
and generates respectively one and two LLRs for each value. 
For 8-PSK, the proposed architecture processes two values (one symbol) to produce three LLRs, using two demapper blocks in parallel for the four values.

The process for the bitwise soft mapper is the inverse of this, using two LLRs per symbol for the QPSK, three for 8-PSK, and four for 16-QAM. The calculation of the EP-based soft estimates is the same for the three constellations, using four of these blocks in parallel.

\begin{table}[h]
\centering
\begin{tabular}{|c|c|c|c|c|c|c|c|c|}
\hline
\multicolumn{2}{|c}{XC7Z020-} & \multicolumn{2}{|c|}{LUTs} & \multicolumn{2}{|c|}{Flip-Flops} & \multicolumn{2}{|c|}{BRAMs} & \multirow{2}{3.7em}{Critical Path (ns)} \\\cline{3-8}
\multicolumn{2}{|c|}{1CLG400C} & \ftextnumero & \% & \ftextnumero & \% & \ftextnumero & \% & \\\hline
\multicolumn{2}{|c}{Resources} & \multicolumn{2}{|c}{53200} & \multicolumn{2}{|c|}{106400} & \multicolumn{2}{|c|}{140} & - \\\hline
\multirow{2}{3em}{QPSK} & A & 1035 & 1.95 & 198 & 0.19 & 0.5 & 0.36 & \multirow{2}{3.7em}{9.418} \\\cline{2-8}
 & B & 4204 & 7.90 & 5373 & 5.05 & 2.5 & 1.79 & \\\hline
\multirow{2}{3em}{8-PSK} & A & 1805 & 3.39 & 346 & 0.33 & 0.5 & 0.36 & \multirow{2}{3.7em}{9.648} \\\cline{2-8}
 & B & 4968 & 9.34 & 5521 & 5.19 & 2.5 & 1.79 & \\\hline
\multirow{2}{3em}{16-QAM} & A & 1448 & 2.72 & 340 & 0.32 & 0.5 & 0.36 & \multirow{2}{3.7em}{9.738} \\\cline{2-8}
 & B & 4654 & 8.75 & 5515 & 5.18 & 2.5 & 1.79 & \\\hline
\multicolumn{9}{l}{A: FD-SILE without HIL configuration.} \\\multicolumn{9}{l}{B: FD-SILE with HIL configuration.}
\end{tabular}
\caption{FPGA Resources Table.}
\label{tab_resources}
\end{table}


The FPGA implementation demonstrates efficient resource usage, as shown in Table \ref{tab_resources}, with very limited occupation in terms of Look-Up Tables (LUTs) and Flip-Flops (FFs). The implementation of the analytical blocks consumes fewer resources than other resources that are necessary to communicate with the ARM processor. It is important to note that the implementation with the highest resource usage is the one for 8-PSK mapping. Indeed, this mapping uses a LUT-aided semi-analytical implementation instead of an analytical implementation. One can note that no DSP resources are assigned in the FPGA. This is due to the fact that all the arithmetic operations are applied on data represented by a limited number of bits, thanks to the proposed fixed-point analysis. Furthermore, all implementations achieve an execution frequency of 100MHz, given the critical paths below 10 ns.

\section{Conclusion}
\label{Concl}

This paper has presented a comprehensive study and, to the best of our knowledge, the first hardware implementation of an EP-based FD-SILE for QPSK, 8-PSK, and 16-QAM constellations on an FPGA platform. 

The results demonstrated that the fixed-point version performed comparably to the floating-point one, even helping to mitigate the negative impacts of bad interference patterns in the challenging Proakis-C test channel.
The resource utilization for the analytical blocks was low, allowing the remaining resources to be used for other blocks of the receiver if necessary.


In future works, there are several promising possibilities. One such direction is the implementation of a flexible receiver that can adapt to the mapping constellation. Another possibility is the integration of turbo iterations into the receiver design, which could further enhance its performance.

\bibliographystyle{unsrt}

\vspace{12pt}
\color{red}

\newpage

\end{document}